\documentstyle[12pt,a4,epsfig]{article}
\newcommand{\be}{\begin{equation}}
\newcommand{\ee}{\end{equation}}
\newcommand{\ea}{\end{eqnarray}}
\newcommand{\baa}{\begin{eqnarray*}}
\newcommand{\eaa}{\end{eqnarray*}}
\newcommand{\bb}{\begin{thebibliography}}
\newcommand{\eb}{\end{thebibliography}}

\newcommand{\bi}[1]{\bibitem{#1}}
\begin {document}

\hfill DESY 00-016

\hfill hep-ph/0002091
\vspace{2cm}

\begin{center}
{\Large \bf Single-spin Azimuthal Asymmetries \\ 
in the ``Reduced Twist-3 Approximation''}\\

\vspace*{1cm}
E.~De~Sanctis$^a$, W.-D.~Nowak$^{b}$, K.A.~Oganessyan$^{a,b,c}$
\footnote{e-mail: kogan@hermes.desy.de} \\
\vspace*{0.3cm}
{$^a$\it INFN-Laboratori Nazionali di Frascati\\
I-00044 Frascati, via Enrico Fermi 40, Italy}\\
\vspace*{0.3cm}
{$^b$\it DESY Zeuthen\\ 
D-15738 Zeuthen, Platanenallee 6, Germany}\\
\vspace*{0.3cm}
{$^c$\it Yerevan Physics Institute\\
375036 Yerevan, Alikhanian Br.2, Armenia}\\
\end{center}

\vspace{2cm}

\begin{abstract}
We consider the single-spin azimuthal asymmetries recently measured at 
the HERMES experiment for charged pions produced in  
semi-inclusive deep inelastic scattering of leptons off longitudinally 
polarized protons. Guided by the experimental results 
and assuming a vanishing twist-2 {\it transverse} quark spin distribution
in the {\it longitudinally} polarized nucleon, denoted  as ``reduced 
twist-3 approximation'', a self-consistent 
description of the observed single-spin asymmetries is obtained. In addition, 
predictions are given for the $z$ dependence of the single target-spin 
asymmetry. 
\end{abstract}

\newpage

Semi-inclusive deep inelastic scattering (SIDIS) of leptons off a polarized 
nucleon target is a rich source of information on the spin structure of the 
nucleon and on parton fragmentation. In particular, 
measurements of azimuthal asymmetries in SIDIS allow the further 
investigation of the quark and gluon structure of the polarized nucleon. 
The HERMES collaboration
has recently reported on the measurement of single target-spin asymmetries 
in the distribution of the azimuthal angle $\phi$ relative to the lepton 
scattering plane, in semi-inclusive charged pion production on a 
longitudinally polarized hydrogen target~\cite{HERM}. The $\sin\phi$ 
moment of this distribution was found to be 
significant for $\pi^{+}$-production. For $\pi^{-}$ it was found to 
be consistent with zero within present experimental uncertainties, as it 
was the case for the $\sin2\phi$ moments of both $\pi^{+}$ and $\pi^{-}$.
Single-spin asymmetries vanish in models in which hadrons consist of 
non-interacting collinear partons (quarks 
and gluons), i.e. they are forbidden in the simplest version of the 
parton model and perturbative QCD. 
Non-vanishing and non-identical intrinsic transverse 
momentum distributions for oppositely polarized partons play an important 
role in most explanations of such non-zero single-spin asymmetries; they 
are interpreted as the effects of ``naive time-reversal-odd'' (T-odd) 
fragmentation functions [2-6], arising 
from non-perturbative hadronic final-state interactions. In Refs.~\cite{KO,EP} 
these asymmetries were evaluated and it was shown that a good agreement 
with the HERMES data can be achieved by using only twist-2 
distribution and fragmentation functions.  \\

In this letter the single target-spin {\it sin}$\,\phi_h$ and 
{\it sin}$\,2\phi_h$ 
azimuthal asymmetries are investigated in the light of the recent HERMES 
results~\cite{HERM}. It will be shown that these results may be interpreted
towards a vanishing twist-2 quark transverse spin distribution in the 
longitudinally  polarized nucleon~\footnote{After this work has been completed we became aware of 
Refs.~\cite{BM0,EFR0} where this possibility has also been considered.}.
Under this assumption, which will be called hereafter ``reduced twist-3 
approximation'', the sub-leading order in $1/Q$ single target-spin 
{\it sin}$\,\phi_h$ asymmetry reduces to the twist-2 level and is 
interpreted as the effect of the convolution of the transversity 
distribution and the T-odd fragmentation function. In this situation, 
also measurements with a longitudinally polarized target at HERMES
may be used to extract the transversity distribution in a way 
similar to that proposed in Ref.~\cite{KNO} for a transversely polarized 
target, once enough statistics will be collected. \\

The {\it sin}$\,\phi_h$ and {\it sin}$\,2\phi_h$ moments of experimentally 
observable single target-spin asymmetries in the SIDIS cross-section can be 
related to the parton distribution and fragmentation functions involved in
the parton level description of the underlying process \cite{AK,TM}. Their
anticipated dependence on $p_T$ ($k_T$), the intrinsic transverse momentum of 
the initial (final) parton, reflects into the distribution of $P_{hT}$, the 
transverse momentum of the semi-inclusively measured hadron.
The moments are defined as appropriately weighted integrals over this 
observable, of the cross section asymmetry:

\begin{equation}
\langle \frac{{\vert P_{hT}\vert}}{M_h} \sin \phi_h \rangle \equiv 
\frac{\int d^2P_{hT} \frac{{\vert P_{hT}\vert}}{M_h}
\sin \phi_h \left(d\sigma^{+}-d\sigma^{-}\right)}
{\int d^2P_{hT} \left(d\sigma^{+} + d\sigma^{-}\right)},
\end{equation}
\begin{equation}
\langle \frac{{\vert P_{hT}\vert}^2}{MM_h} \sin 2\phi_h \rangle \equiv 
\frac{\int d^2P_{hT} \frac{{\vert P_{hT}\vert}^2}{MM_h}
\sin 2\phi_h \left(d\sigma^{+}-d\sigma^{-}\right)}
{\int d^2P_{hT} \left(d\sigma^{+} + d\sigma^{-}\right)}.
\end{equation}
Here $+ (-)$ denote the antiparallel (parallel) longitudinal polarization of 
the target and $M$ ($M_h$) is the mass of the target (final hadron). 
For both polarized and unpolarized
leptons these asymmetries are given by~\cite{AK,TM,OABK}~\footnote{We omit 
the current quark mass dependent terms.}
\begin{equation}
\langle \frac{\vert P_{hT}\vert}{M_h} \sin \phi_h \rangle (x,y,z) = 
{1 \over I_0(x,y,z)} [I_{1L}(x,y,z) + I_{1T}(x,y,z)], 
\label{AS}
\end{equation}
\begin{equation}
\langle \frac{{\vert P_{hT}\vert}^2}{MM_h} \sin 2\phi_h \rangle (x,y,z) = 
{8 \over I_0(x,y,z)} S_L (1-y) \, h_{1L}^{\perp(1)}(x)
z^2 H_1^{\perp (1)}(z), 
\label{AS1}
\end{equation}
where 
$$
I_0 (x,y,z) = (1+(1+y)^2) f_1(x) D_1(z),
$$
\begin{equation}
I_{1L} (x,y,z) = 4S_L {M \over Q}\,(2-y)\sqrt{1-y} \, 
[ x h_L(x) z H_1^{\perp (1)}(z) - h^{\perp(1)}_{1L}(x) \tilde{H}(z) ],
\label{ASL}
\end{equation}
\begin{equation}
I_{1T} (x,y,z) = 2 S_{T\,x}\,(1-y)\,h_1(x) z H_1^{\perp (1)}(z). 
\label{AST}
\end{equation}
With $k_1$ ($k_2$) being the 4-momentum of the incoming (outgoing) charged lepton, 
$Q^2=-q^2$ where $q=k_1-k_2$ is the 4-momentum of the virtual photon. $P$ 
($P_h$) is the momentum of the target (final hadron), $x=Q^2/2(Pq)$, 
$y=(Pq)/(Pk_1)$, $z=(PP_h)/(Pq)$, $k_{1T}$ the incoming lepton
transverse momentum with respect to the virtual photon momentum direction, and 
$\phi_h$ is the azimuthal angle between $P_{hT}$ and $k_{1T}$ around the 
virtual photon direction. Note that the azimuthal angle of the transverse 
(with respect to the virtual photon) component of the target polarization, 
$\phi_S$, is equal to 0 ($\pi$) for the target polarized parallel 
(anti-parallel) to the beam \cite{OABK}. The components of the longitudinal 
and transverse target polarization 
in the virtual photon frame are denoted by $S_L$ and $S_{Tx}$, respectively.
Twist-2 distribution and fragmentation functions have a subscript `1':
$f_1(x)$ and $D_1(z)$ are the usual unpolarized distribution and 
fragmentation functions, while 
$h^{\perp (1)}_{1L}(x)$ and $h_1(x)$ describe the quark transverse spin 
distribution in longitudinally and transversely polarized nucleons,  
respectively. The twist-3 distribution function in the longitudinally 
polarized nucleon is denoted by $h_L(x)$ ~\cite{JJ91}.  
The spin dependent fragmentation function $H_1^{\perp(1)}(z)$, describing 
transversely polarized quark fragmentation (Collins effect~\cite{COL}), 
can be interpreted as the production probability of an unpolarized 
hadron from a transversely polarized quark~\cite{MUL}. The fragmentation 
function $\tilde{H}(z)$ is the interaction-dependent part of the twist-3 
fragmentation function: $H(z)=-2zH_1^{\perp(1)}(z)+\tilde{H}(z) $. 
The functions with superscript $(1)$ denote $p^2_T$- and $k^2_T$-moments,
respectively: 
\be
h_{1L}^{\perp (1)}(x) \equiv \int d^2p_T\,{\left(\frac{p_T^2}{2M^2}
\right)}\, h_{1L}^{\perp}(x, p_T^2),
\label{WF1} 
\ee  
\be
H_1^{\perp (1)}(z) \equiv z^2 \int d^2k_T\,{\left(\frac{k_T^2}{2M^2_h}
\right)} H_1^{\perp}(z, z^2 k_T^2).  
\label{WF2} 
\ee  
The function $h_L(x)$ can be split into a twist-2 part, $h^{\perp (1)}_{1L}(x)$,  
and an interaction-dependent part, $\tilde{h}_L(x)$~\cite{JJ91,MT}: 
\be
h_L(x) = -2{h^{\perp(1)}_{1L}(x) \over x}+\tilde{h}_L(x).
\label{HL1}  
\ee
As it was shown in Refs.~\cite{MT,TM} this relation can be rewritten as 
\be
h_L(x) = h_1(x)- {d \over dx}h^{\perp(1)}_{1L}(x).
\label{HL2} 
\ee

The weighted single target-spin asymmetries defined above are related to 
the ones measured by HERMES~\cite{HERM} through the following relations:  
\begin{equation}
 A^{\sin\phi_h}_{UL} 
\approx {2M_h \over {\langle P_{hT}\rangle }} \langle \frac{\vert 
P_{hT}\vert}{M_h} \sin \phi_h \rangle, 
\label{REL1}
\end{equation}
\begin{equation}
A^{\sin2\phi_h}_{UL} 
\approx {2MM_h \over {\langle P^2_{hT}\rangle }} \langle \frac{\vert 
P^2_{hT}\vert}{MM_h} \sin 2\phi_h \rangle, 
\label{REL2}
\end{equation}
where the subscripts $U$ and $L$ indicate unpolarized beam and longitudinally 
polarized target, respectively.  

When combining the HERMES experimental results of a 
significant target-spin {\it sin}$\,\phi_h$ asymmetry for $\pi^{+}$ and 
of a vanishing {\it sin}$\,2\phi_h$ asymmetry
with the preliminary evidence from $Z^0 \to $ $2$-jet 
decay on a non-zero T-odd transversely polarized quark fragmentation 
function~\cite{EFR}, it follows immediately from Eq.(\ref{AS1}) that 
$h^{\perp(1)}_{1L}(x)$, the twist-2 transverse quark spin distribution in 
a longitudinally polarized nucleon, should vanish. Consequently, from Eqs. 
(\ref{HL1}, \ref{HL2}) follows that
\be
h_L(x) = \tilde{h}_L(x) = h_1(x).
\label{HL} 
\ee  
In this situation the single target-spin {\it sin}$\,\phi_h$ asymmetry 
given by Eq.(\ref{AS}) reduces to the twist-2 level (``reduced 
twist-3 approximation''). The fact that 
$h^{\perp(1)}_{1L}(x)$ (see Eq.\ref{WF1}) vanishes may be interpreted as 
follows: the 
distribution function $h^{\perp}_{1L}(x,p_T^2)$, which is non-zero 
itself, vanishes at any $x$ when it is averaged over the intrinsic transverse 
momentum of the initial parton, $p_T$. As a matter of fact, in a 
longitudinally polarized nucleon partons polarized 
transversely at large $p_T$ may 
indeed have a polarization opposite to that at smaller $p_T$, at any $x$. 

It is important to mention that the ``reduced twist-3 approximation'' 
does not require $\tilde{H}(z)=0$, which otherwise would lead to 
the inconsistency that $H^{\perp(1)}_1(z)$ would be required to 
vanish~\cite{TM,ST}.       \\

For the numerical calculations the non-relativistic approximation $h_1(x) = 
g_1(x)$ is taken as lower limit~\footnote{For non-relativistic quarks 
$h_1(x) = g_1(x)$. Several models suggest that  
$h_1(x)$ has the same 
order of magnitude as $g_1$~\cite{JJ92,PP,BCD}. The evolution properties 
of $h_1$ and 
$g_1$, however, are very different~\cite{AM,SV}. At the $Q^2$ values of the 
HERMES measurement 
the assumption $h_1=g_1$ is fulfilled at large, i.e. valence-like, $x$ values, 
while large differences occur at lower $x$~\cite{KNO}. }, 
and $h_1(x)=(f_1(x)+g_1(x))/2$ as an upper limit~\cite{SOF}.  
For the sake of simplicity, $Q^2$-independent parameterizations were chosen  
for the distribution functions $f_1(x)$ and $g_1(x)$~\cite{BBS}. 

To calculate the T-odd fragmentation function $H_1^{\perp (1)}(z)$ ,
the Collins parameterization~\cite{COL} for the analyzing power of 
transversely polarized quark fragmentation was adopted:
\begin{equation}
A_C(z,k_T) \equiv \frac{\vert k_T \vert}{M_h}\frac{H_1^{\perp}(z,k_T^2)}
{D_1(z,k_T^2)} = \frac{M_C\,\vert k_T \vert}{M_C^2+k_T^2}
\label{H1T}
\end{equation}

\vspace{-1cm}

\begin{figure}[ht]
\begin{center}
\epsfig{file=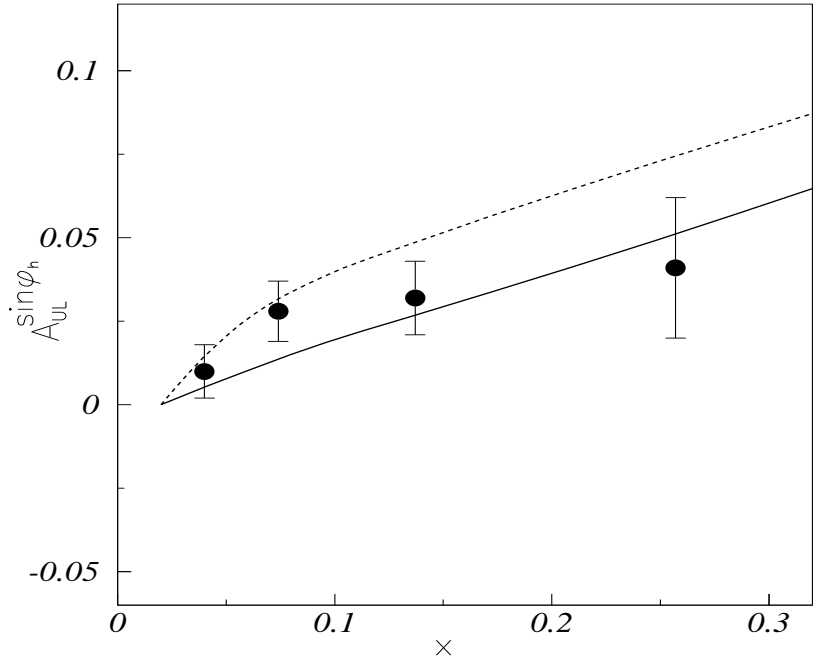, height=10cm,width=12cm}
\caption{The single target-spin asymmetry $A^{\sin\phi_h}_{UL}$ for $\pi^{+}$ 
production as a function of Bjorken $x$, evaluated using $M_C=0.28$ GeV 
in Eq.(\ref{H1T}). The solid line corresponds to $h_1=g_1$, 
the dashed one to $h_1=(f_1+g_1)/2$. Data are from Ref.~\cite{HERM}. } 
\label{fig:f1}
\end{center}
\end{figure}

For the distribution of the final parton's intrinsic transverse momentum, 
$k_T$, in the unpolarized fragmentation function $D_1(z,k^2_T)$ a Gaussian 
parameterization was used~\cite{KM} with $\langle z^2 k_T^2 \rangle = b^2$ 
(in the numerical calculations $b = 0.36$ GeV was taken~\cite{PYTHIA}). 
For $D_1^{\pi^{+}} (z)$ the parameterization from Ref.~\cite{REYA} was adopted.
In Eq.(\ref{H1T}) $M_C$ is a typical hadronic mass whose value may range from
$m_{\pi}$ to $M_p$. Using $M_C=2m_{\pi}$ for the analyzing power of 
Eq.(\ref{H1T}) results in    
\begin{equation}
\frac{\int_{z_0=0.1}^{1} dz H^{\perp}_1(z)}{\int_{z_0=0.1}^{1} dz D_1(z)} 
= 0.062, 
\label{rat}
\end{equation}
which is in good agreement with the experimental result $0.063\pm 0.017 $ 
given for this 
ratio in Ref.~\cite{EFR}. Here $H_1^{\perp}(z)$ is the unweighted polarized 
fragmentation function, defined as: 
\be
H_1^{\perp}(z) \equiv z^2 \int d^2k_T\, H_1^{\perp}(z, z^2 k_T^2).  
\label{WF0} 
\ee  
It is worth mentioning  
that the ratio in Eq.({\ref{rat}}) is rather 
sensitive to the lower limit of integration, $z_0$~\cite{BM}. By  
using $z_0 =0.01$, the ratio  
reduces to 0.03; choosing a value of $z_0$   
equal to 0.2 (0.3), the ratio increases to about $0.1$ ($0.12$). This 
behavior is mainly due to the fact that the fragmentation 
function $D_1(z)$ diverges at small values of $z$.   \\

In Fig.~\ref{fig:f1}, the asymmetry $A^{\sin\phi_h}_{UL}(x)$ of Eq.(\ref{REL1})
for $\pi^{+}$ production on a proton target is presented as a function of
$x$-Bjorken and compared to HERMES data~\cite{HERM}, which   
correspond to $1$ GeV$^2$ $\leq Q^2 \leq 15$ GeV$^2$, $4$ GeV $\leq E_{\pi} 
\leq 13.5$ GeV, $0.02 \leq x \leq 0.4$, $0.2 \leq z \leq 0.7$, and $0.2 
\leq y \leq 0.8$. The two theoretical curves are calculated by integrating 
over the same kinematic ranges taking $\langle P_{hT} \rangle = 0.365$ GeV 
as input. The latter value is obtained in this
kinematic region assuming a Gaussian parameterization of 
the distribution and fragmentation functions 
with $\langle p_T^2 \rangle=(0.44)^2$ GeV$^2$~\cite{PYTHIA}. 

From Fig.~1 it can be concluded that there is good agreement 
between the calculation in this letter and the HERMES data. 
Note that the `kinematic` contribution to $A^{\sin\phi_h}_{UL}(x)$, coming
from the transverse component of the target polarization (with respect to
the virtual photon direction) and given by 
$I_{1T}$ (Eq.(\ref{AST})), amounts to only $25\%$. 

\vspace{-0.5cm}

\begin{figure}[ht]
\begin{center}
\epsfig{file=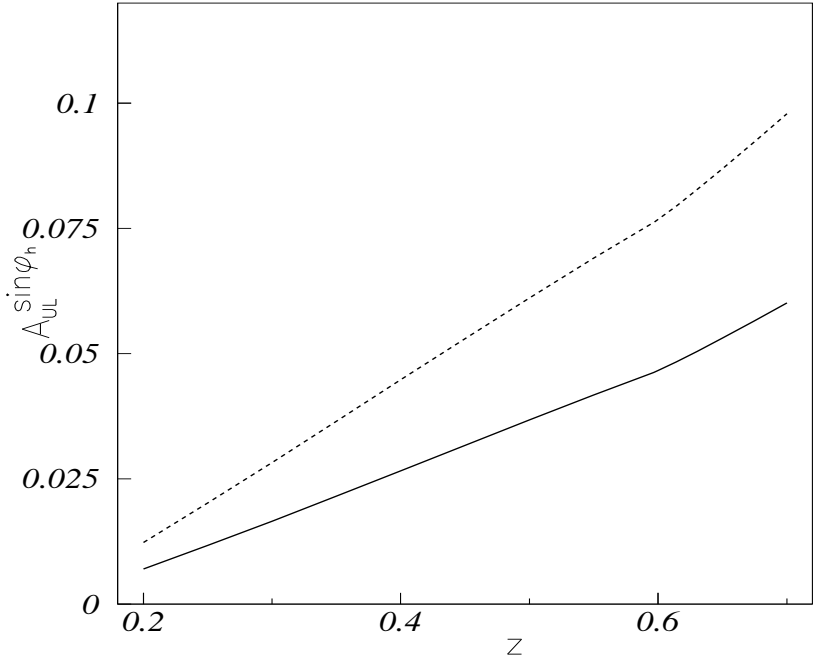, height=10cm,width=12cm}
\caption{The single target-spin asymmetry $A^{\sin\phi_h}_{UL}$ for $\pi^{+}$ 
production as a function of $z$ evaluated using $M_C=0.28$ GeV.  
The solid line corresponds to $h_1=g_1$, the dashed one to $h_1=(f_1+g_1)/2$.}
\label{fig:f2}
\end{center}
\end{figure}

The $z$ dependence of the asymmetry $A^{\sin\phi_h}_{UL}$ for $\pi^{+}$ 
production is shown in Fig.~\ref{fig:f2}, where  
the two curves correspond to two limits for $h_1(x)$, 
as introduced above. No data are available yet to constrain the
calculations. \\   

In conclusion, the recently observed single-spin azimuthal asymmetries 
in semi-inclusive deep inelastic lepton scattering off a longitudinally 
polarized proton target at HERMES are interpreted on the basis of the 
so-called ``reduced twist-3 approximation'', that is assuming a vanishing 
twist-2 transverse quark spin distribution in the longitudinally 
polarized nucleon. This leads to a 
self-consistent description of the observed single-spin asymmetries. 
In this approach the target-spin {\it sin}$\,\phi_h$ asymmetry is 
interpreted as the effect of the convolution of the transversity 
distribution, $h_1(x)$, and a T-odd fragmentation 
function, $H^{\perp(1)}_{1}(z)$, and may allow to probe 
transverse spin observables in a longitudinally polarized nucleon. 

In addition, predictions are given for the $z$ dependence of the single 
target-spin {\it sin}$\,\phi_h$ asymmetry, for which experimental data 
are not yet published.   

\vspace*{0.7cm} 

We would like to thank P.~Mulders for many useful discussions, V.~Korotkov 
for very useful comments and 
R.~Kaiser for the careful reading of the manuscript.
The work of K.A.O. was in part supported by INTAS
contributions (contract numbers 93-1827 and 96-287) from the European Union.

\bb{99}

  \bi{HERM} HERMES Collaboration, A. Airapetian, et.al., hep-ex/9910062. 
  \bi{COL} J.~Collins, Nucl. Phys. B {\bf 396} (1993) 161.
  \bi{AK} A.~Kotzinian, Nucl. Phys. B {\bf 441} (1995) 234.
  \bi{TM} P.J.~Mulders and R.D.~Tangerman, Nucl Phys. B {\bf 461} (1996) 197. 
  \bi{ARTRU} X.~Artru, J.~Czyzevski and H.~Yabuki, Z. Phys C {\bf 73}, 
               (1997) 527. 
  \bi{JAF1} R.~Jaffe, X.~Ji and J.~Tang, Phys. Rev. Lett. {\bf 80} (1998) 1166. 
  \bi{KO} A.M.~Kotzinian, K.A.~Oganessyan, A.R.~Avakian, E.~DeSanctis, 
                hep-ph/9908446;  Proceedings of the workshop Nucleon'99, Frascati, 
                 June 7-9, 1999, Nucl. Phys. {\bf A666}.  
  \bi{EP} A.V.~Efremov, M.~Polyakov, K.~Goeke, D.~Urbano, hep-ph/0001119. 
   \bi{BM0}  M. Boglione and P.J. Mulders, hep-ph/0001196.  
  \bi{EFR0} A.V.~Efremov, hep-ph/0001214. 
  \bi{KNO}  V.A.~Korotkov, W.-D.~Nowak, K.A.~Oganessyan, DESY 99-176,
            hep-ph/0002268. 
  \bi{OABK} K.A.~Oganessyan, A.R.~Avakian, N.~Bianchi, A.M.~Kotzinian, 
                hep-ph/9808368;  Proceedings of the workshop Baryons'98, Bonn, 
                 Sept. 22-26, 1998.
  \bi{JJ92} R.~Jaffe, X.~Ji, Nucl. Phys. {\bf B375} (1992) 527.
  \bi{JJ91} R.~Jaffe, X.~Ji, Phys. Rev. Lett, {\bf 67} (1991) 552. 
  \bi{MUL}  P.J.~Mulders, Nucl. Phys. {\bf A622} (1997) 239c. 
  \bi{MT} R.D.~Tangerman, P.J.~Mulders, NIKHEF-94-P7, hep-ph/9408305. 
  \bi{EFR} A.V.~Efremov, O.G.~Smirnova and L.G.~Tkachev,
                hep-ph/9812522; A.V.~Efremov, Proceedings of workshop
                DIS'99, Zeuthen, 19-23 April 1999. 
  \bi{ST} A.~Schaefer, O.V.~Teryaev, hep-ph/9908412. 
  \bi{PP} P.V.~Pobylitsa and M.V.~Polyakov, Phys. Lett. {\bf B389} (1996) 350.  
  \bi{BCD} V.~Barone, T.~Calarco and A.~Drago, Phys. Lett. {\bf B390} (1997) 287.
  \bi{AM} X~.Artru, M~.Mekhfi, Zeit. Phys. {\bf C45}, 669 (1990)
  \bi{SV} S.~Scopetta, V.~Vento, Phys.Lett. {\bf B424} (1998) 25.  
  \bi{SOF} J.~Soffer, Phys. Rev. Lett. {\bf 74} (1995) 1292.
  \bi{BBS} S.~Brodsky, M.~Burkardt, I.~Schmidt, 
                 Nucl. Phys. {\bf B441} (1995) 197. 
  \bi{KM} A.M.~Kotzinian, P.J.~Mulders, Phys. Lett. {\bf B406} (1997) 373.  
  \bi{PYTHIA} T.~Sjostrand, Comp. Phys. Commun. {\bf 82} (1994) 74; CERN-TH.7112/93; 
               hep-ph/9508391. 
  \bi{REYA} E.~Reya, Phys. Rep. {\bf 69} (1981) 195. 
  \bi{BM} E. Boglione, P.J.~Mulders, Phys. Rev. {\bf D60} (1999) 054007. 

\eb

\end{document}